\documentclass[twocolumn]{revtex4}
\usepackage{amssymb} %maths
\usepackage{amsmath} %maths
\usepackage{graphicx}

\begin{document}
%opening
\title{Thermo-refractive and thermo-chemical noise in the beamsplitter of GEO600 gravitational-wave interferometer.}
\author{Bruin Benthem}
\author{Yuri Levin}
\affiliation{Leiden University, Leiden Observatory and Lorentz Institute, 
Niels Bohrweg 2, 2300 RA Leiden, the Netherlands}

\begin{abstract}
Braginsky, Gorodetsky, and Vyatchanin have shown that  thermo-refractive 
fluctuations are an important source of noise in interferometric gravitational-wave detectors.
In particular, the thermo-refractive noise in the GEO600 beamsplitter is expected to make a 
substantial contribution to the interferometer's total noise budget. 
Here we present a new computation of the GEO600 thermo-refractive  noise which takes 
into account the beam's elliptical profile and, more importantly, the fact that the laser beam 
induces a standing electromagnetic wave in the beamsplitter.
The use of updated parameters results in the overall reduction of the calculated noise amplitude by
a factor of $\sim 5$ in the low-frequency part of the GEO600 band, compared to the 
previous estimates.  
We also find, by contrast with previous calculations, that thermo-refractive fluctuations
result  in white noise between 600 Hz and 39 MHz, at a level of $8.5\cdot 10^{-24}$Hz$^{-1/2}$. 
Finally, we describe a new type of thermal noise, which we call the thermo-chemical noise.
This is caused by a random motion of optically-active chemical impurities or structural defects
 in the direction along a steep intensity gradient of the standing wave. 
We discuss the potential relevance of the thermo-chemical noise for GEO600.
\end{abstract}

\maketitle
\section{Introduction and main results for thermo-refractive noise} 
The optical layout of GEO600, the German-British gravitational-wave detector, differs in an essential
way from that of LIGO and VIRGO, the two other working gravitational-wave interferometers.
In both LIGO and VIRGO each of the arms consists of a high-quality Fabry-Perot cavity, where
most of the light power is concentrated, while the beamsplitter is located outside of both cavities.
By contrast, in GEO600 the light, after passing the beamsplitter, is processed only once through each of
the doubly-folded arms \cite{wilke}. Therefore, GEO600 is much more sensitive to the noise originating
at the beamsplitter than its LIGO/VIRGO counterparts.

The beamsplitter in the  GEO600 laser interferometer induces a difference in the optical path lengths 
between the interferometer's two arms since one of the beams is transmitted through the beamsplitter 
twice while is the other is being reflected from its surface. 
This difference in the optical path length can be modified by a change in the refractive index of the beamsplitter 
due to random fluctuations of temperature in the refracting material. 
As Braginsky, Gorodetsky, and Vyatchanin have shown 
\cite{BGV}, the temperature fluctuations result from a random heat flow between different parts of the refractive
medium. 
Since the interrogating beam is not homogeneous but has a Gaussian profile 
the inhomogeneous temperature fluctuations produce a net (fluctuating) change in the transmitted beam's phase 
which is seen as noise in the overall output of the instrument. For a circular Gaussian
beam, Braginsky and Vyatchanin \cite{BV} have computed the following spectral density of
this thermo-refractive noise expressed as an equivalent change in the end mirror's position $\delta z$:
\begin{equation} 
S^{\delta z}(\omega) = \frac{4k_B\kappa T^2\beta^2 a}{\pi(C\rho r_0^2 \omega)^2}
\label{BVresult}
\end{equation}
Here, $k_B$ is Boltzmann's constant, $\kappa$ the thermal conductivity, $T$ the temperature, $\beta = \partial n/\partial T$ where 
$n$ is the refractive index, $a$ is the beamsplitter's thickness, $C$ the specific heat, $\rho$ the density and $r_0$ is the beam's 
radius defined in terms of the beam intensity, so that 
$I=I_0\exp(-r^2/r_0^2)$. The expression in Eq.~(\ref{BVresult}) has been used to estimate the thermo-refractive noise
in the GEO600 beamsplitter \cite{Reid}. 

However, inside the GEO600 beamsplitter, the beam is not circular: 
the $45^\circ$ angle of incidence and the refractive index $n=1.45$ of fused silica introduces ellipticity of the beam. 
The ratio of the major and minor axes of the beam's elliptical crossection
is given by $\eta=1.23$, while the minor axis equals the width of the original beam. 
More importantly, the beam is not homogeneous in the longitudinal direction
since a standing wave is formed in the beamsplitter as a result of the beam's reflection from the end mirror.
The changes in the material's refractive index \cite{footnote1} close to the nodes 
of the electric-field intensity 
have less effect on the wave's phase than changes in the antinodes. 
Therefore, random thermal fluctuations on the scale of the light wavelength are important, 
making for a much higher level of noise at frequencies higher than $\sim 600$ Hz than the BV result would suggest. 

\begin{figure}[!h]
\begin{center}
\includegraphics[width = 0.4\textwidth]{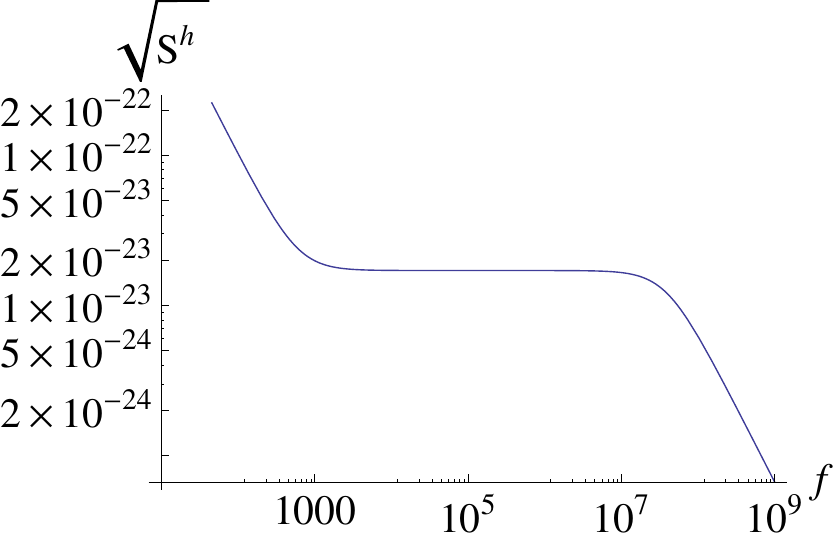}
\end{center}
\caption{loglogplot of the noise amplitude $\sqrt{S^h(f)}$ in units of Hz$^{-1/2}$ as a function of $f$ in units of Hz with the GEO-600 parameters}\label{plot}
\end{figure}

\begin{figure*}[h]
\begin{center}
\includegraphics[width = 0.9\textwidth]{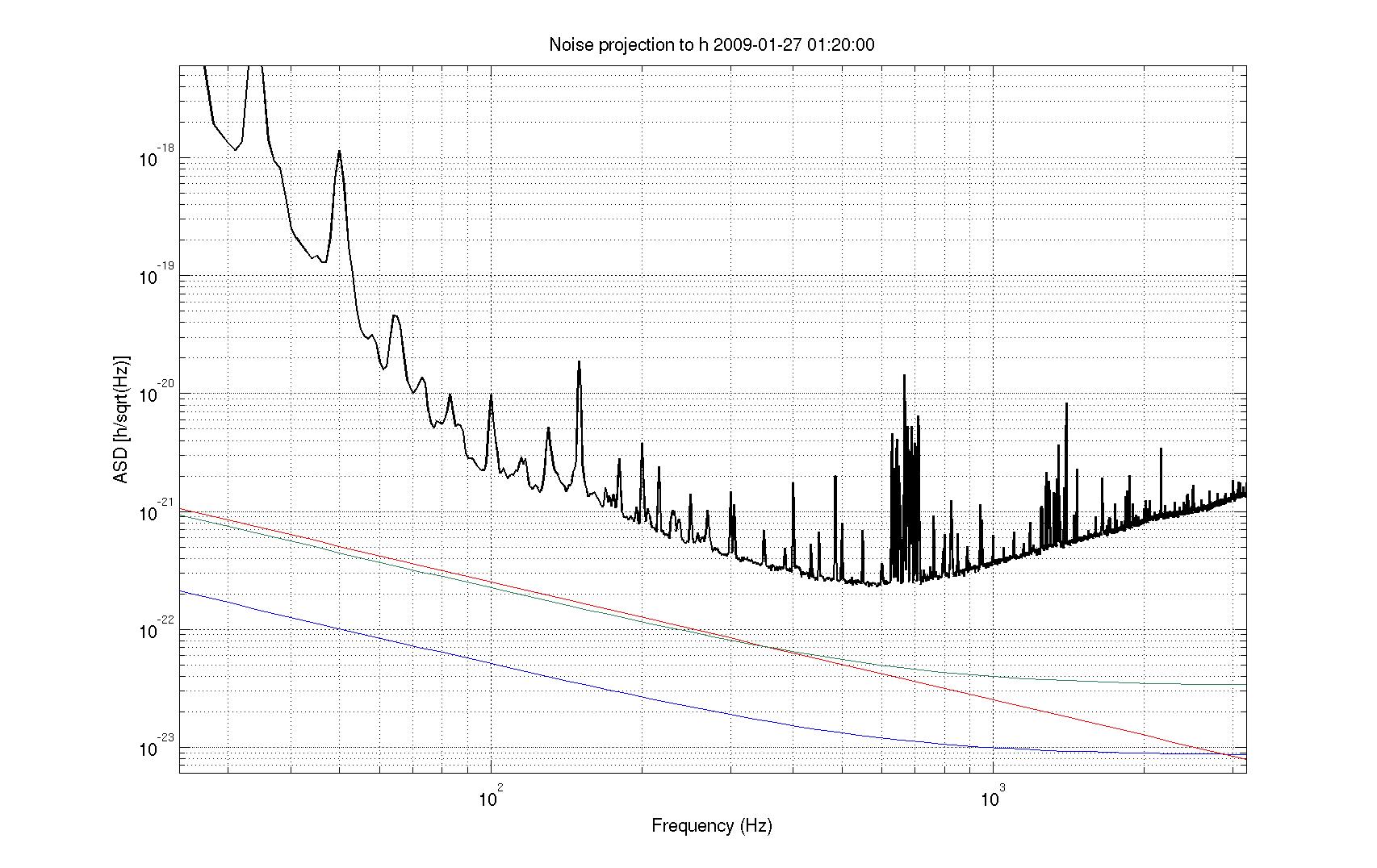}
\end{center}
\caption{Plot of measured GEO600 noise (black line), beamsplitter thermo-refractive noise from previous calculations
\cite{Reid} (red line) and corrected  thermo-refractive noise using the same parameters (green line), and finally of 
corrected thermo-refractive noise using the updated parameters from \cite{Jerome} (blue line,
\cite{footnote3})}\label{plot2}
\end{figure*}

In this paper we derive the expression for the thermo-refractive noise with these corrections
taken into account. 
We obtain the following result:
\begin{align}
S^{\delta z}(\omega) = &\frac{4k_B\kappa T^2\beta^2a'}{\pi(C\rho r_0^2\omega)^2}\frac{\eta + \eta^{-1}}{2\eta^2}\cdot \nonumber \\ 
&\left[1+\frac{2k^2r_0^2\eta}{(\eta+\eta^{-1})(1+(2kl_{th})^4)}\right] 
\label{result}
\end{align}
Here $a'=a/\cos(i)$ is the optical pathlength through the beamsplitter, where $i=\arcsin(1/\sqrt{2}n)\sim 29^\circ$ is the angle 
between the beam axis and the normal to the beamsplitter, 
$k=2\pi n/\lambda$ is the wavevector and $l_{th} = \sqrt{\kappa/(C\rho\omega)}$ is the thermal diffusion length. 
The numerical values relevant for GEO600 are given in table \ref{table1}. 
\begin{table}[!h]
\begin{center}
\begin{tabular}{|c|c|} 
\hline
Symbol & (GEO600-)Value \\
\hline\hline
$k_B$ & $1.38 \cdot 10^{-23}$ J/K \\
\hline
$\kappa$ & 1.38 W/mK \\
\hline
$T$ & $300$ K \\
\hline
$\beta$ & $8.5 \cdot 10^{-6}$\\
\hline
$a$ & $8.0\cdot 10^{-2}$m\\
\hline
$\eta$ & 1.23 \\
\hline
$k$ & $8.56 \cdot 10^6$ m$^{-1}$ \\
\hline
$r_0$ & $0.71\cdot 10^{-2}$m \\
\hline
$C$ & 746 J/kgK \\
\hline
$\rho$ & 2200 kg/m$^3$ \\
\hline
$L$ & 1200 m \\
\hline
\end{tabular}
\end{center}
\caption{Parameter values taken from \cite{Jerome}}\label{table1}
\end{table}
The first term on the right-hand side of Eq.~(\ref{result})  is the contribution 
due to the transverse elliptic profile of the beam, and it 
agrees with the result in \cite{BV} for $\eta=1$, while the second term is due 
to the presence of the standing wave. 
It is common to express the noise amplitude in terms of the dimensionless metric $h$ for easy comparison to other sources. 
Figure \ref{plot} is a plot of $\sqrt{S^h} = \sqrt{S^{\delta z}}/L$ where $L$ is the interferometers {\it unfolded} 
armlength and figure \ref{plot2} 
is a comparison to the measured noise spectrum and the previous estimate for the beamsplitter thermo-refractive noise \cite{footnote3}.
 

At a typical gravitational wave frequency of 100 Hz the noise amplitude is $\sqrt{S^h} \sim 5.1\cdot 10^{-23}$Hz$^{-1/2}$. 
For low frequencies the effect of the standing wave is negligible and 
we have the $1/f$ 
dependence, consistent  with the previous calculations. However, for higher frequencies 
the standing-wave contribution takes over and the noise spectrum becomes white between $600$Hz and $39$Mhz 
with an amplitude $\sqrt{S^h} \sim 8.5\cdot 10^{-24}$Hz$^{-1/2}$. 
At the latter frequency the thermal diffusion length becomes comparable to the wavelength of the beam light,
and the $1/f$ dependence is recovered but at much higher value than would be predicted by the  BV formula.

\section{Method of calculation}

The foundation of this analysis is direct method  \cite{Levin} 
for calculating  thermo-refractive noise.  
The relevant readout variable is the  phase of the output beam, 
translated into a change in the end mirror's position $\delta z$. 
It is a fairly straightforward calculation to show 
\begin{equation} \delta z = \int_V d^3\vec{r} \cdot a' \cdot \beta \cdot \delta T(\vec{r}) \cdot q(\vec{r}), 
\label{genvar}
\end{equation} 
where the integral is over the beamsplitter's volume $V$, $\delta T(\vec{r})$ is the temperature fluctuation 
at $\vec{r}$, and $q(\vec{r})$ is the form-factor given, taking the elliptic Gaussian beam 
and standing wave into account, by 
\begin{equation} q(\vec{r}) = \frac{2}{\pi r_0^2 a' \eta} \exp\left[-\left(\frac{x^2}{r_0^2}+\frac{y^2}{(\eta r_0)^2}\right)
\right]\sin^2(kz).
\label{formfac}
\end{equation}

We now calculate the noise of this generalized variable via the hypothetical
experiment proceeding  in three steps:
\begin{enumerate}
 \item Periodically inject entropy into the medium with volume density 
\begin{equation} \frac{\delta S(\vec{r},t)}{dV} = F_0\cos(\omega t) a'\beta q(\vec{r})
\end{equation}
\item Calculate the dissipated power $W_\text{diss}$ as a result of this entropy injection. 

To do the latter, we solve the heat equation 
\begin{equation} 
C\rho\frac{\partial \delta T}{\partial t} - \kappa \nabla^2\delta T 
= T\frac{\partial}{\partial t}\frac{\delta S(\vec{r},t)}{dV}
\end{equation} 
while keeping in mind that: 
\begin{itemize} 
\item the solution is periodic with frequency $\omega$ 
\item the wavelength is much smaller than the beamsplitter thickness so we can ignore boundary effects 
\item the diffusion of the osscilating temperature in the transverse direction is negligible, giving 
$\nabla^2 = \partial^2/\partial z^2$, since $r_0 >> l_{th}$ for all frequencies of interest. 
\end{itemize} 

The dissipated power is then computed by using the standard expression 
\begin{equation} 
W_\text{diss}=\int_V d^3\vec{r} \frac{\kappa}{T}\langle(\nabla \delta T)^2\rangle 
\end{equation} 
taken from e.g.~Landau and Lifshitz \cite{LL}, where $\langle ...\rangle$ denotes the time average over one period.
\item Calculate the spectral density of the noise in $\delta z$ using the fluctuation dissipation theorem 
\cite{Levin}, \cite{CW}:
\begin{equation} 
S^{\delta z}(\omega) = \frac{8k_B T}{\omega^2}\frac{W_\text{diss}}{F_0^2}\label{fdt}\end{equation}
\end{enumerate}
These three steps lead directly to the main result in Eq.~(\ref{result}).

\section{Thermo-chemical noise}

The small scale of the wavelength opens up the possibility for a new type of thermal noise which has not yet been considered. 
The fused silica used for contemporary beamsplitters contains minute quantities 
of contaminants such as OH ions, Cl ions and other defects which have an effect on the refractive 
index depending on their concentration. As these optically-active contaminants diffuse up and 
down the steep gradient of the standing-wave electric-field intensity, they  cause fluctuations of the overall 
beam's phaseshift. 

Lets compute this thermo-chemical noise. Let $P(\vec{r})$ be the fluctuating volume
concentration of the optically active impurities. 
The optical path change due to these impurities is given by
\begin{equation} 
\delta z = \int_V d^3\vec{r}\cdot a'\cdot \alpha \cdot \delta P(\vec{r}) \cdot q(\vec{r}),
\end{equation} 
where $\alpha = \partial n/\partial P$ and $q(\vec{r})$ is given by Eq.~\ref{formfac}.

We can follow our earlier treatment of thermal noise \cite{Levin2} and 
calculate the dissipated power in the system under the Hamiltonian 
\begin{equation} 
H_{\rm int}=- F_0\cos(\omega t)\delta z
\end{equation} 
This can be easily done by recalling that the formal expression for $P(\vec{r})$ is
\begin{equation}
P(\vec{r})=\Sigma_i \delta(\vec{r}-\vec{r}_i),
\label{P}
\end{equation}
where $\vec{r}_i$ are the positions of individual optical impurities.
Then, under the action of the above Hamiltonian,  each impurity experiences a force
\begin{equation}
f_i=F_0\cos(\omega t) a' \alpha {\partial q(\vec{r_i})\over\partial z_i}.
\end{equation}
Under the action of this force, the impurity drifts and energy is dissipated.
We assume that the impurities achieve 
their terminal drift velocity $v_i$ on a timescale much shorter than the frequencies of interest, and
use Einstein's relation
\cite{Einstein} 
\begin{equation} 
Df_i = v_ik_BT
\end{equation} 
to compute the drift velocity \cite{footnote2}.
Here $D$ is the diffusion co\"efficient. The dissipated power per particle is $<f_iv_i>$;
summing over the particles and substituting the result into Eq.~(\ref{fdt}) yields 
\begin{equation} S^{\delta z}(\omega) = \frac{4DP\alpha^2k^2a'}{\pi r_0^2 \eta \omega^2}\label{result2}\end{equation}

Taking values of Suprasil 311 SV used in GEO-600 for OH ions \cite{Takke}, 
$P(OH) = 3.9\cdot 10^{24} \text{m}^{-3}$ and $\alpha = -4.52\cdot 10^{-31} \text{m}^3$ and 
estimating a value for $D$ from \cite{Diffusion} to be in the order 
of $10^{-20}\text{m}^2/\text{s}$ at room temperature we get 
\begin{equation} 
\sqrt{S^h(\omega)} = 4.4 \cdot 10^{-26}\text{Hz}^{-1/2}
\end{equation} 
at a frequency of 100 Hz, suggesting that a simple version of the thermo-chemical
noise is outside the realm of relevance for the GEO600 interferometer. 
However, there may be other optically-active mobile impurities (e.g., small structural defects 
or localized 2-state systems) in glass that have not yet been considered.
Potential presence of such impurities must be thoroughly investigated in future work.

\section{Discussion}
The calculations presented here demonstrate that the thermo-refractive fluctuations result in a
noise floor which, while substantially below currently measured noise,
may become important for future updates of the GEO600 interferometer. The standing-wave contribution
should be taken into account in all future calculations of the thermo-refractive noise from transmissive
optics.

As a side product, we have identified a new type of thermal noise: the thermo-chemical noise, which is also enabled by the
presence of the optical standing wave in the beamsplitter. The naive estimates of this noise, which
are based on what is known about the chemical impurities in Suprasil 311, place it beyond
the realm of concern. However, one needs to be vigilant about other types of mobile impurities which have
not yet been identified, and which could potentially produce a higher thermo-chemical noise.

\section{Acknowledgements}
We thank Jerome Degallaix for bringing our attention to thermo-refractive noise in GEO600 beamsplitters,
and for authoritative updates on the current GEO600 parameters.
We thank Stuart Reid for showing us his preprint with the thermo-refracrive noise estimates based 
on the Braginsky-Vyatchanin formula, and Kip Thorne and Wim van Saarloos for useful discussions. 
We also thank Ralf Takke of Heraeus for providing information about chemical impurities
in the Suprasil 311 fused silica glass.


\begin{thebibliography}{99}
\bibitem{footnote1} In the beamsplitter's dielectric material
the change of the index of refraction is caused by the change in the dielectric constant.
\bibitem{footnote3} We remark that there is a significant difference between the updated parameters \cite{Jerome} and 
those used for the calculations done previously \cite{Reid}. Current $\beta$ is $8.5$ ppm compared to the previous
value of $15$ ppm, current $r_0$ is $1/\sqrt{2}$cm compared to the previous $0.88/\sqrt{2}$cm, and it is essencial to
use the length of the unfolded arm $L=1200$m in the calculation of $S_h$ as opposed to the previously used $L=600$m.
All of these corrections, together with the beam ellipticity, reduce the calculated magnitude of the thermo-refractive noise.
\bibitem{footnote2}Another assumption we make is that at frequencies of interest the diffusion
scale is much smaller than the light wavelength. This assumption is satisfied with a very high degree of precision.
\end{thebibliography}
\end{document}